\documentclass[journal=apchd5,manuscript=article]{achemso}

\usepackage[version=3]{mhchem} % Formula subscripts using \ce{}
\usepackage[T1]{fontenc}       % Use modern font encodings
\usepackage[dvipsnames,usenames]{xcolor}
\usepackage{amsmath}
\usepackage{subfig}
\usepackage{tikz}

\author{Matthias Maier}
\affiliation{School of Mathematics, University of Minnesota, Minneapolis, MN 55455, USA}

\author{Andrei Nemilentsau}
\email{anemilen@umn.edu}
\affiliation{Department of Electrical \& Computer Engineering, University of Minnesota, Minneapolis, MN 55455, USA}

\author{Tony Low}
\email{tlow@umn.edu}
\affiliation{Department of Electrical \& Computer Engineering, University of Minnesota, Minneapolis, MN 55455, USA}

\author{Mitchell Luskin}
\affiliation{School of Mathematics, University of Minnesota, Minneapolis, MN 55455, USA}

\title{Ultracompact amplitude modulator by coupling hyperbolic
polaritons over a graphene-covered gap}

\abbreviations{IR,NMR,UV}

\begin{document}


\begin{abstract}
	The hyperbolic phonon-polaritons within the Reststrahlen band of hBN are of great interest for applications in nanophotonics as they are capable of propagating light signals with low losses over large distances. However, due to the phononic nature of the polaritons in hBN, amplitude modulation of its signal proves to be difficult and has been underexplored. In this paper, we propose theoretically a broadband efficient amplitude modulator for hyperbolic rays in hBN operating in the frequency range between 1450 cm$^{-1}$ and 1550 cm$^{-1}$. The modulating region comprises a few tens of nanometers wide gap carved within the hBN slab and covered by a graphene layer, where electrostatically gated graphene serves as a mediator that facilitates the coupling between phonon-polaritons on each side of the gap through plasmonic modes within graphene. We demonstrate that such an ultra compact modulator has insertion losses as low as 3 dB and provides modulation depth varying between 14 and 20 dB within the type-II hyperbolicity region of hBN.

\end{abstract}

Hexagonal boron nitride (hBN) is a natural hyperbolic material that
recently attracted a lot of attention due to its prospective applications
for mid-infrared nano-photonics
\cite{dai2014tunable,caldwell2014sub,caldwell2015low,Dai2015,yoxall2015direct,woessner2015highly,caldwell2016atomic,basov2016polaritons,low2017polaritons}.
The hyperbolic polaritons in hBN within the Reststrahlen band are of
phononic nature and thus experience significantly lower losses
\cite{giles2017ultra} than their plasmonic counterparts in 2D materials
such as graphene, MoS$_2$ and black phosphorus \cite{avouris20172d}. Hence,
hBN based waveguides are promising for efficient channeling of signals over
large distances in the form of highly-confined sub-diffractional rays
\cite{li2015hyperbolic,dai2015subdiffractional}. However, unlike plasmonic
2D materials, where dispersion of plasmon polaritons is defined by the
density of electron gas, the dispersion of phonon-polaritons in hBN,
intrinsically originating from oscillations of lattice atoms in a polar
crystal, is defined predominantly by the crystal structure. Thus, while the
density of electron gas and thus characteristics of polariton modes in
plasmonic materials can be actively tuned by applying an electric
bias\cite{koppens2011graphene,grigorenko2012graphene,low2014graphene,garcia2014graphene,goncalves2016introduction},
achieving a similar degree of control for pure phonon-polariton hBN modes
is difficult. As a way to overcome this issue, a huge swath of research has
been devoted towards coupling phonon-polaritons in hBN to plasmonic modes
in nearby 2D materials, where properties of the hybrid plasmon-phonon modes
can be changed by means of electrostatic gating. This has proven to be an
effective strategy for active control of plasmon-phonon-polariton dispersion in graphene-hBN heterostructures
\cite{Dai2015,Kumar2015,woessner2015highly,barcelos2015graphene,jia2015tunable,yang2016far,caldwell2016atomic}.
An alternative approach towards controlling the polariton characteristics in polar dielectrics relies on utilizing phase change materials (PCMs) \cite{li2016reversible}, where changes in the relative permittivity of the PCMs induced by optical pumping can be exploited to alter the polariton dispersion.

Although electrostatic gating of coupled plasmon-phonon modes provides a
way of controlling their dispersion characteristics, this does not
necessarily lead to an efficient amplitude modulation of a signal carried
by the hybrid modes. Meanwhile, signal modulation is one of the most
crucial operations in nano-photonics and thus utilizing 2D
materials as modulators at telecom wavelengths is of significant interest\cite{sun2016optical}. Particularly, electro-optical modulators
based on graphene have a high speed of operation, with a modulation speed
varying between 1 GHz\cite{liu2011graphene,liu2012double,gao2015high} and
35 GHz \cite{dalir2016athermal} being reported. Moreover, theoretically
imposed limits on the modulation speed exceeds 100
GHz\cite{gosciniak2012theoretical,phare2015graphene}. The main issue with
graphene based optical modulators in near-IR and optical frequency ranges
is the weak intrinsic optical absorption of graphene ($\sim
2.3\%$\cite{mak2008measurement,nair2008fine}) which limits modulation
depths for  single-layer\cite{liu2011graphene,hu2016broadband} and
double-layer\cite{liu2012double} graphene based modulators to about 0.1 dB
$\mu$m$^{-1}$. The strength of light-matter interaction and,
correspondingly, the modulation depth can be improved by integrating
graphene with a dielectric cavity resonator
\cite{gao2015high,phare2015graphene}, with modulation depths as high as 3.2
dB being demonstrated \cite{gao2015high}. Unlike the telecom frequency
range, in the THz frequency range graphene's conductivity can be significantly
modified by applied electric bias and efficient graphene based modulators
have been reported\cite{lee2012switching,sensale2012broadband,sensale2012extraordinary,liang2015integrated}
with up to 47$\%$ modulation being achieved \cite{lee2012switching} for a
modulator operating in the transmission regime. Moreover,
the efficient modulation of a quantum cascade laser gain has been
reported that utilizes plasmons in graphene incorporated
into the laser microcavity for gate
tuning\cite{chakraborty2016gain}. The mid-IR
frequency range provides potential interest for applications in
nano-photonics, as this is the range where polar dielectrics, such as hBN,
support low-loss highly confined phonon-polaritons.
Recently, a nanomechanical graphene/hBN based light
modulator operating in the reflection regime (with modulation depth up to
30 $\%$) has been demonstrated \cite{thomas2016nanomechanical} in the
frequency range from mid-IR up to UV. However, efficient amplitude
modulation of the guided phonon-polaritons in hBN has not been explored
yet.

\begin{figure}[tbhp]
  \centering
  \subfloat[]{\quad\includegraphics[width=0.5\textwidth]{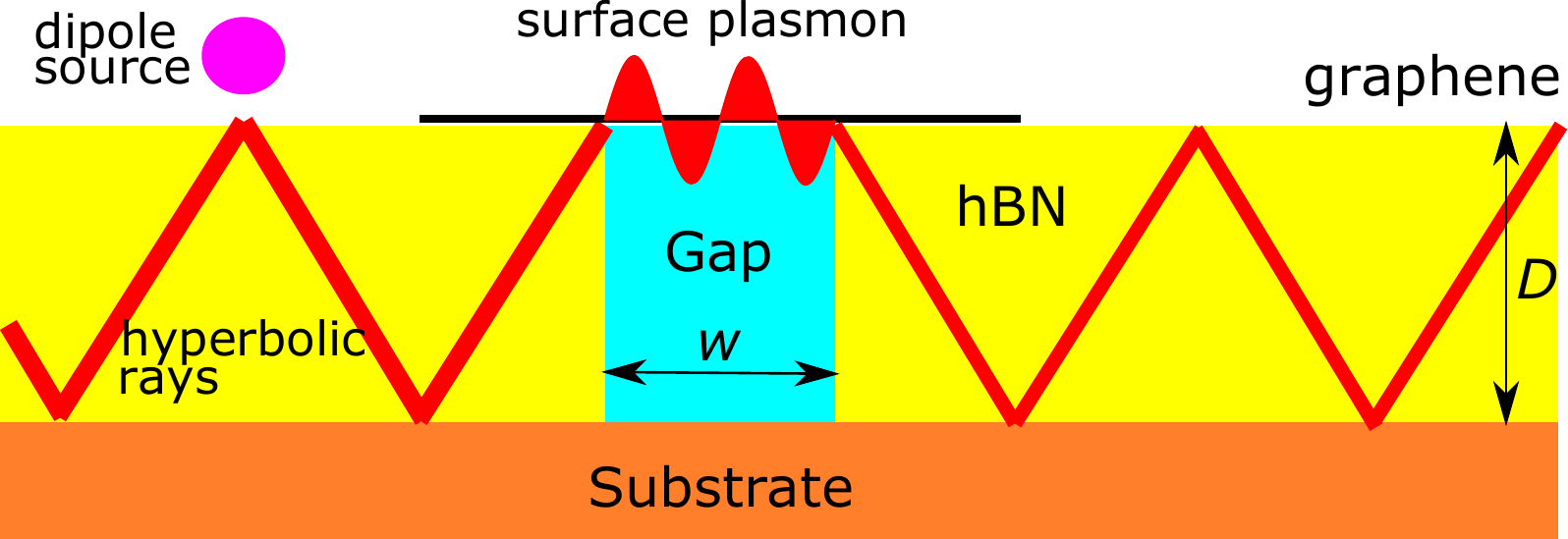}\quad}

  \subfloat[]{\quad\includegraphics[width=0.4\textwidth]{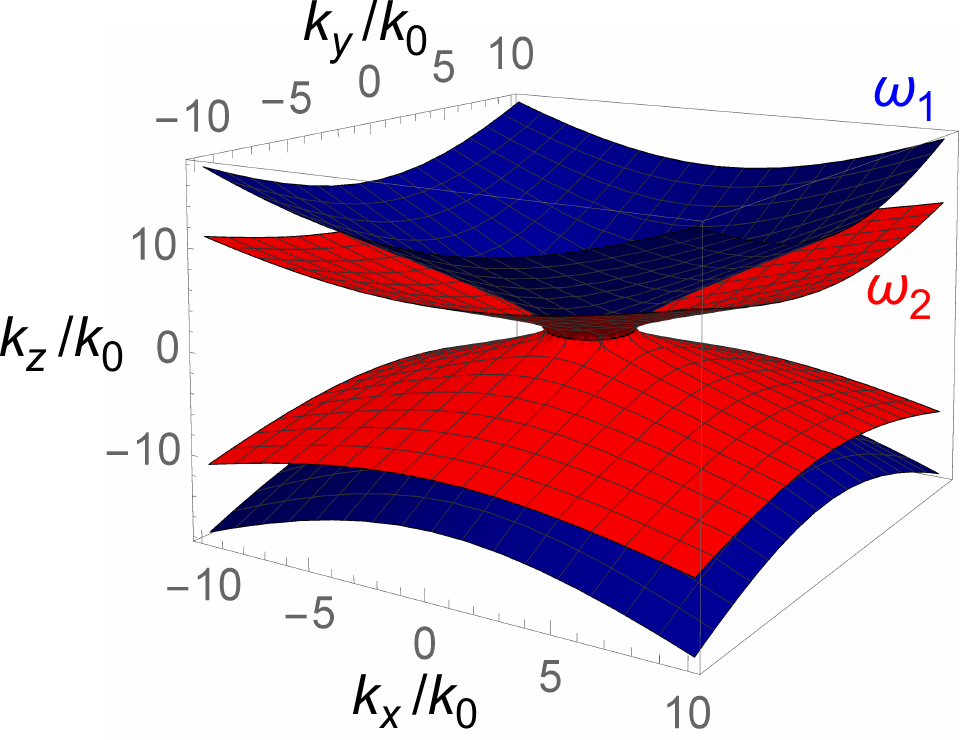}\quad}  \subfloat[]{\quad\includegraphics[width=0.4\textwidth]{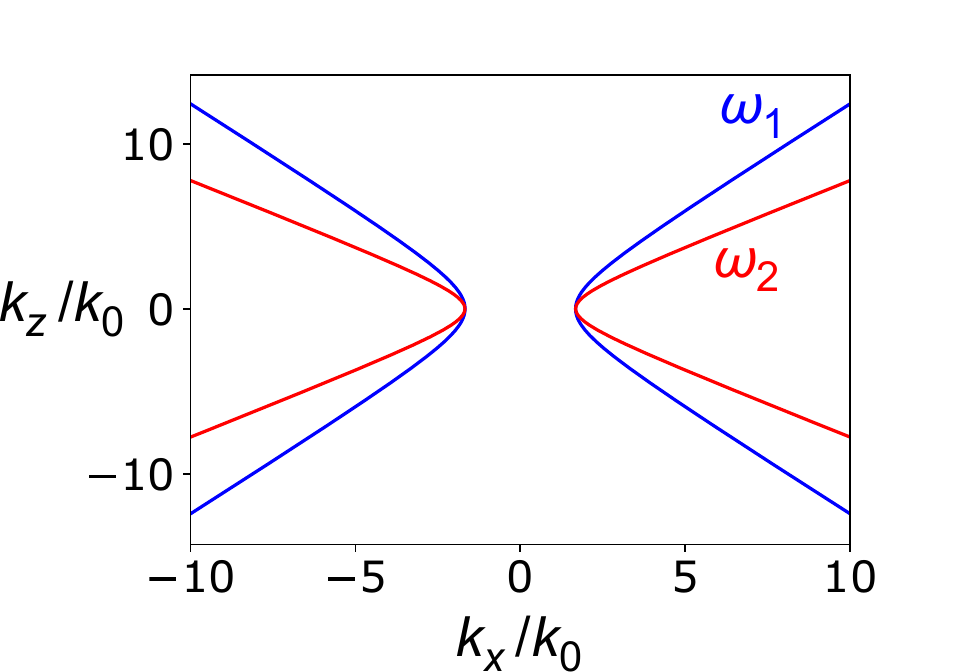}\quad}

  \subfloat[]{\quad\includegraphics[width=0.35\textwidth]{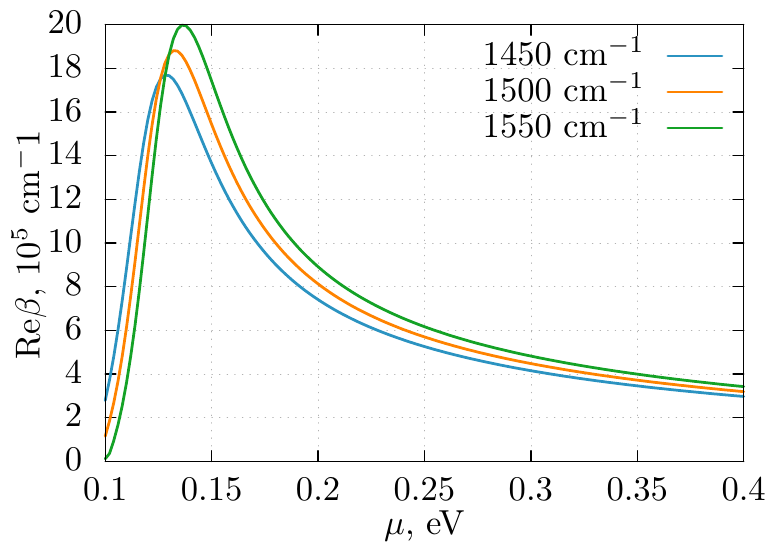}\quad}
  \subfloat[]{\quad\includegraphics[width=0.35\textwidth]{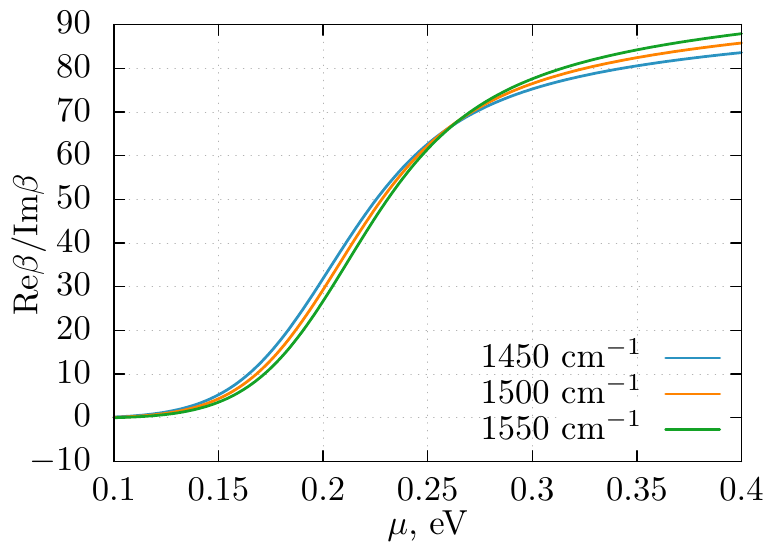}\quad}

  \subfloat[]{\quad\includegraphics[width=0.35\textwidth]{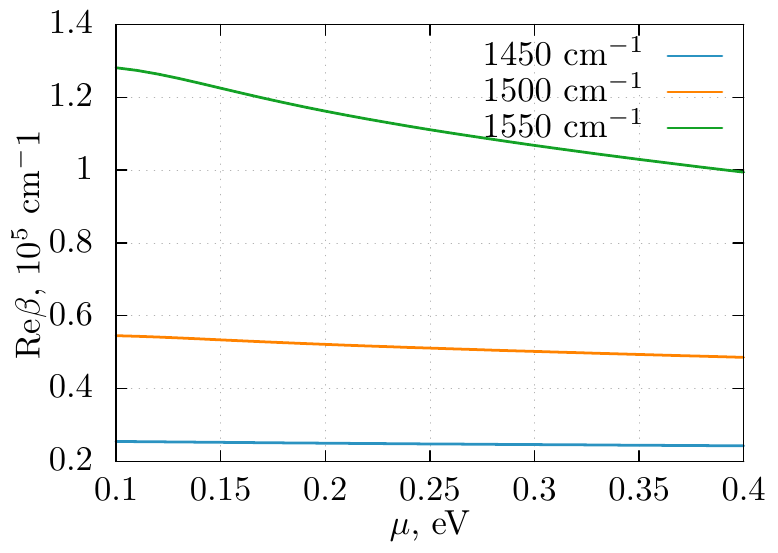}\quad}
  \subfloat[]{\quad\includegraphics[width=0.35\textwidth]{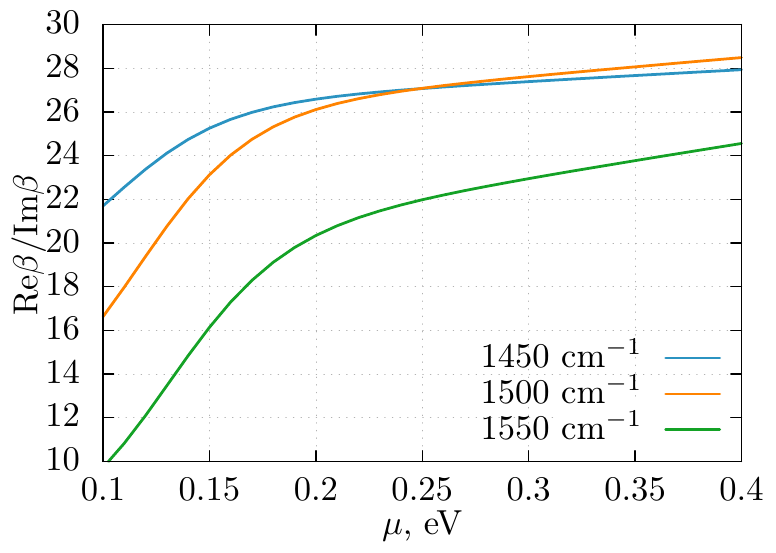}\quad}

  \caption{(a) Sketch of graphene-hBN modulator geometry; (b)
    Isofrequency surface, $\omega(\mathbf{k}) = \mathrm{const}$, and (c)
    its projection on plane $k_y = 0$, for extraordinary wave in hBN at
    frequencies $\omega_1 = 1500$ cm$^{-1}$, $\omega_2 = 1550$ cm$^{-1}$.
    $\mathbf{v}_g = \nabla_{\mathbf{k}} \omega$ is the group velocity in
    the hBN. (d),(e) Wavenumber $\beta$ of a surface plasmon in graphene
    monolayer in vacuum as a function of graphene chemical potential $\mu$
    for three different frequencies indicated in panel (d). Relaxation time
    $\tau = 0.34$ps and graphene temperature $T = 300$K. (f,g) Wavenumber
    $\beta$ of a first mode in air-graphene-hBN-substrate system, where $D =
    100$ nm and relative permittivity of substrate $\varepsilon_s = 1.5$.}
    \label{Figure1}
\end{figure}

In this paper, we demonstrate that efficient modulation of hyperbolic
phonon-polaritons in hBN is indeed possible. We propose an ultra-compact
graphene-hBN plasmonic amplitude modulator with a modulation depth of up to
20 dB achieved within a modulating region less than 100 nm long in the
frequency range 1450-1550 cm$^{-1}$, in conjunction with insertion losses
of only 3 dB. A sketch of the modulator geometry is presented in
Fig.~\ref{Figure1}a. The core of the modulator is a graphene-covered hBN
slab, which, in the frequency range under consideration, allows for signals
to propagate in a form of highly directional hyperbolic rays with low
losses over large distances. For simplicity, we assume that the signal is
injected into the modulator by a line-current source. The modulating region
comprises a gap filled with a dielectric and covered by a graphene layer
that mediates the coupling between the hyperbolic modes in two neighboring
hBN regions. When the graphene is biased to a chemical potential of 0.22 eV, we
observe high ($\sim$ 50$\%$) transmission of the hyperbolic mode across
the gap. However, the transmission falls more than an order of magnitude
when the graphene bias is reduced to 0.12 eV. This can be explained by the
strong plasmonic losses within graphene in this range of chemical
potentials.

In order to elaborate on the physics underlying the modulator performance, we
need to clarify the nature of elementary excitations in the system. In what
follows, we restrict our consideration to the frequency range 1400-1570
cm$^{-1}$, i.e. the window of type-II hyperbolicity in hBN
materials\cite{Kumar2015,Dai2015}. In this frequency range the optical
response of the system is dominated by hyperbolic phonon-polaritons in hBN,
plasmons in graphene and hybridized polaritons at the hBN-graphene
interface.

The core of the modulator system is an hBN slab of height $D$ that allows
for signals to propagate over long distances with small losses. The dielectric
function of hBN is that of an uniaxial crystal\cite{Kumar2015} with the
crystal optical axis orthogonal to the hBN-graphene interface, i.e.
$\hat{\varepsilon} = diag\{\varepsilon_{\perp}, \varepsilon_{\perp},
\varepsilon_{\parallel}\}$, where $\varepsilon_{\parallel} \approx 2.8$ is
the component of permittivity tensor parallel to  the optical axis, and
\begin{equation}
  \varepsilon_{\perp}(\omega) = \varepsilon_{\infty} + \varepsilon_{\infty}
  \frac{\omega_{\mathrm{LO}}^2 - \omega_{\mathrm{TO}}^2
  }{\omega_{\mathrm{TO}}^2 - \omega^2 - i\omega\gamma}
\end{equation}
is the component of permittivity tensor orthogonal to the optical axis, where
$\varepsilon_{\infty} = 4.8$, $\omega_{\mathrm{TO}} = 1370$ cm$^{-1}$,
$\omega_{\mathrm{LO}} = 1610$ cm$^{-1}$ and $\gamma = 5$ cm$^{-1}$. In
particular, $\varepsilon_{\perp} = -10.55 + i0.5$, $-4.46 + i0.19$, and
$-1.76 + i 0.097$ at $\omega = 1450$ cm$^{-1}$, 1500 cm$^{-1}$, and 1550
cm$^{-1}$, respectively. As one can see $\varepsilon_{\parallel} \cdot
\mathrm{Re} \, \varepsilon_{\perp} < 0$, which indicates that the hBN behaves
as hyperbolic material in the above-mentioned frequency range. The term
"hyperbolic" stems from the fact that the isofrequency surfaces for an extraordinary wave in lossless materials\cite{Poddubny2013} ($\mathrm{Im}\,\varepsilon_{\perp} = 0)$, which are given by
\begin{equation} \label{Eq:equifreq}
  \frac{k_x^2 + k_y^2}{\varepsilon_{\parallel}} +
  \frac{k_z^2}{\mathrm{Re}\,\varepsilon_{\perp}} = k_0^2,
\end{equation}
take the form of a hyperboloid (see Figs.~\ref{Figure1}b, c).  Although hBN is lossy, the losses are extremely small ($\mathrm{Im}\,\varepsilon_{\perp} \ll \mathrm{Re}\,\varepsilon_{\perp}$). Thus, for all practical purposes, hBN behaves as an effectively hyperbolic material.
The profile is further reduced to a cone for large values of photon momenta
$|k_x|, |k_y|, |k_z| \gg k_0$, where $k_0 = \omega/c$ is the free-space
wavenumber, $\omega$ is the angular frequency, and $c$ is the speed of
light in vacuum. The direction of energy flow in anisotropic media is
defined by the group velocity\cite{Kong1986}, $\mathbf{v}_g =
\nabla_{\mathbf{k}} (\omega(\mathbf{k}))$, which points in the direction
normal to the isofrequency surface $\omega(\mathbf{k}) = \mathrm{const}$.
This means that for sufficiently large values of photon momenta the
hyperbolic mode carries energy in a particular direction, constituting an
angle $\theta$ ($\tan \theta = \sqrt{|\mathrm{Re}\,\varepsilon_{\perp}| /
\varepsilon_{\parallel}}$) with respect to the optical axis, which is
entirely defined by the dielectric function of hBN. These modes are
schematically shown as narrow red lines in Fig.~\ref{Figure1}a. Although
high directionality and low losses allow hyperbolic polaritons in hBN to
carry energy over long distances, their phononic nature prevents them from
being responsive to electric tuning and thus pure hBN hyperbolic modes
cannot be used for signal modulation.

The tunability issue can be resolved by coupling hyperbolic
phonon-polaritons in hBN with surface plasmons in graphene as shown in
Fig.~\ref{Figure1}a. The wavenumber $\beta$ of a surface plasmon in a
graphene monolayer sandwiched between two half-spaces with relative
permittivities $\varepsilon_1$ and $\varepsilon_2$ is given by the
non-linear equation \cite{Hanson2008}
\begin{equation} \label{Eq:disper_TM}
  \varepsilon_1 \gamma_2 + \varepsilon_2 \gamma_1 + \frac{i \eta_0 \sigma
  }{c} \gamma_1 \gamma_2 = 0,
\end{equation}
where $\eta_0 = 377$ $\Omega$ is the vacuum impedance, $\gamma_i = \sqrt{\beta^2 -
k_0^2 \varepsilon_i}$, and
\begin{align}
  \label{Eq:graphene_conductivity}
  \sigma(\omega) & = i \frac{e^2 k_B T}{\pi\hbar^2 (\omega + i/\tau)}
  \left(\frac{\mu}{k_B T} + 2 \ln\left(e^{-\mu/k_B T} + 1\right)\right)
  \notag
  \\
  & + \frac{i e^2 (\omega + i/\tau)}{\pi \hbar^2} \int_0^{\infty}
  \frac{f_d(-\epsilon) - f_d(\epsilon)}{(\omega + i/\tau)^2 - 4
  (\epsilon/\hbar)^2} d\epsilon,
\end{align}
is the optical graphene conductivity \cite{Falkovsky2007,Hanson2008}.
Here, $e$ denotes the charge of an electron, $\hbar$ is the reduced
Plank's constant, $k_B$ is the Boltzmann constant, $\mu$ is the graphene
chemical potential, $f_d(\epsilon) = (1 + e^{(\epsilon-\mu)/k_B T})^{-1}$
is the Fermi-Dirac distribution, $\tau = 0.34$ ps and $T=300$ K are
electron relaxation time and temperature, respectively. The dispersion and
quality factor for surface plasmons in graphene layers in vacuum (i.e.
$\varepsilon_1 = \varepsilon_2 = 1$) are shown in Figs.~\ref{Figure1}d, e.
The graphene plasmons are indeed highly tunable, with graphene losses
decreasing significantly as the chemical potential of graphene increases
from 0.1 to 0.4 eV and electron mobility decreases from 34000 cm$^2$V/s to 8500 cm$^2$V/s (see SI for relation between mobility and chemical potential in graphene). These mobility values are reasonable for suspended graphene. \cite{bolotin2008ultrahigh}.

The problem, however, arises from the fact that coupled graphene-hBN
surface modes are predominantly of phonon character in the frequency range
of interest and thus respond significantly more weakly to changes of the
graphene chemical potential than graphene plasmons do. The surface modes in
air-graphene-hBN-substrate system (see  Fig.~\ref{Figure1}a) are defined by
the non-linear equation (see SI for derivation)
\begin{align}
  \label{Eq:surface_hBN-graphene}
  1 - \frac{ ( 1 - \zeta_0\alpha_0) (1-\alpha_s) }{( 1 + \zeta_0 \alpha_0)
  (1 + \alpha_s)} e^{-2\gamma_h D} = 0,
\end{align}
where
$\alpha_0 = \gamma_h /\gamma_0\varepsilon_{\perp}$,
$\alpha_s = \gamma_h \varepsilon_s/\gamma_s \varepsilon_{\perp}$,
$\zeta_0 = 1 - \sigma \eta_0 \gamma_0/i c$,
$\gamma_h = \sqrt{\beta^2 (\varepsilon_{\perp}/\varepsilon_{\parallel}) - k_0^2 \varepsilon_{\perp}}$,
$\gamma_0 = \sqrt{\beta^2 - k_0^2}$, $\gamma_s = \sqrt{\beta^2 - k_0^2 \varepsilon_s}$,
and $\varepsilon_s$ is the relative permittivity of the substrate. For a
given frequency $\omega$, the non-linear equation has an infinite number of
solutions indicating that such a material slab supports an infinite number
of surface modes with increasing wavenumber $\beta$. The dependence of the
wavenumber for the first hybridized graphene-hBN surface mode (i.e. mode
with the smallest wavenumber at a given frequency) on the chemical
potential of graphene is presented in Figs.~\ref{Figure1}f, g for the case
of a hBN slab of thickness $D = 100$ nm and a relative permittivity
$\varepsilon_s = 1.5$ for the substrate. We are mostly concerned with the
first phonon mode as this mode couples most strongly to the dipole
excitation and is observed most prominently in scanning near-field optical
microscopy and scattering experiments
\cite{caldwell2013low,dai2014tunable,Dai2015}. It should be noted, that
dispersion of higher order modes is qualitatively similar to that of the
first-order mode (see SI for details).

The mode wavenumber increases significantly with frequency ranging from
1450 to 1550 cm$^{-1}$. This can be explained by the decrease of
electrostatic screening with smaller hBN permittivity
($\varepsilon_{\perp}$ decrease from -10 to -2). However, at any given
frequency, dependence of the mode wavenumber (and thus losses) on chemical
potential remains weak and thus a bare graphene-hBN system cannot be used to
create an efficient optical or polaritonic modulator.

\begin{figure}[tbhp]
  \centering
  \subfloat[]{\quad\includegraphics[width=0.42\textwidth]{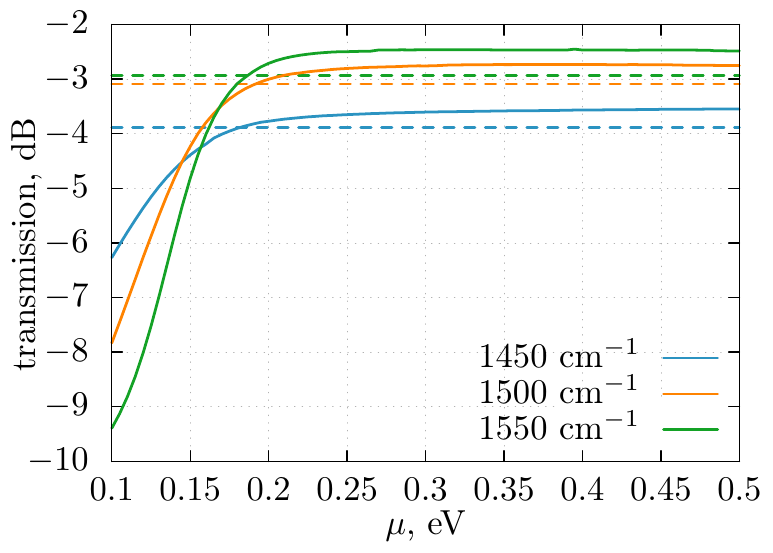}\quad}
  \subfloat[]{\quad\includegraphics[width=0.42\textwidth]{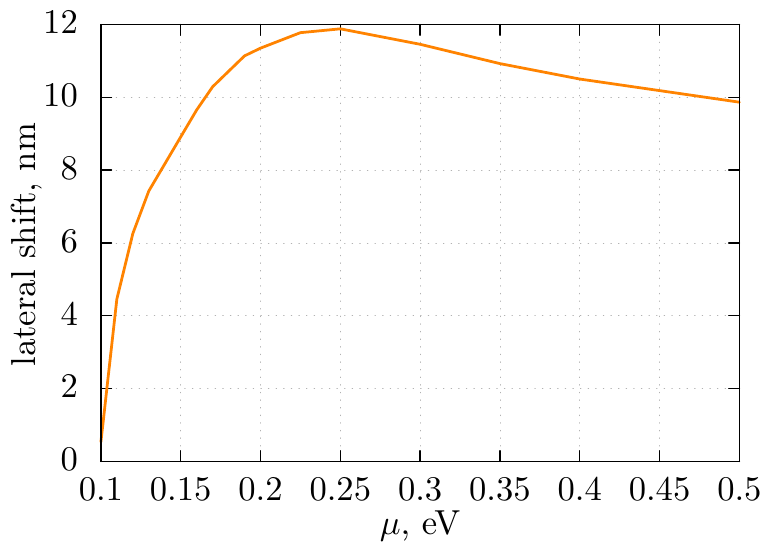}\quad}

  \subfloat[]{\quad
  \begin{tikzpicture}
    \node at (0,0)
      {\includegraphics[width=0.42\textwidth]{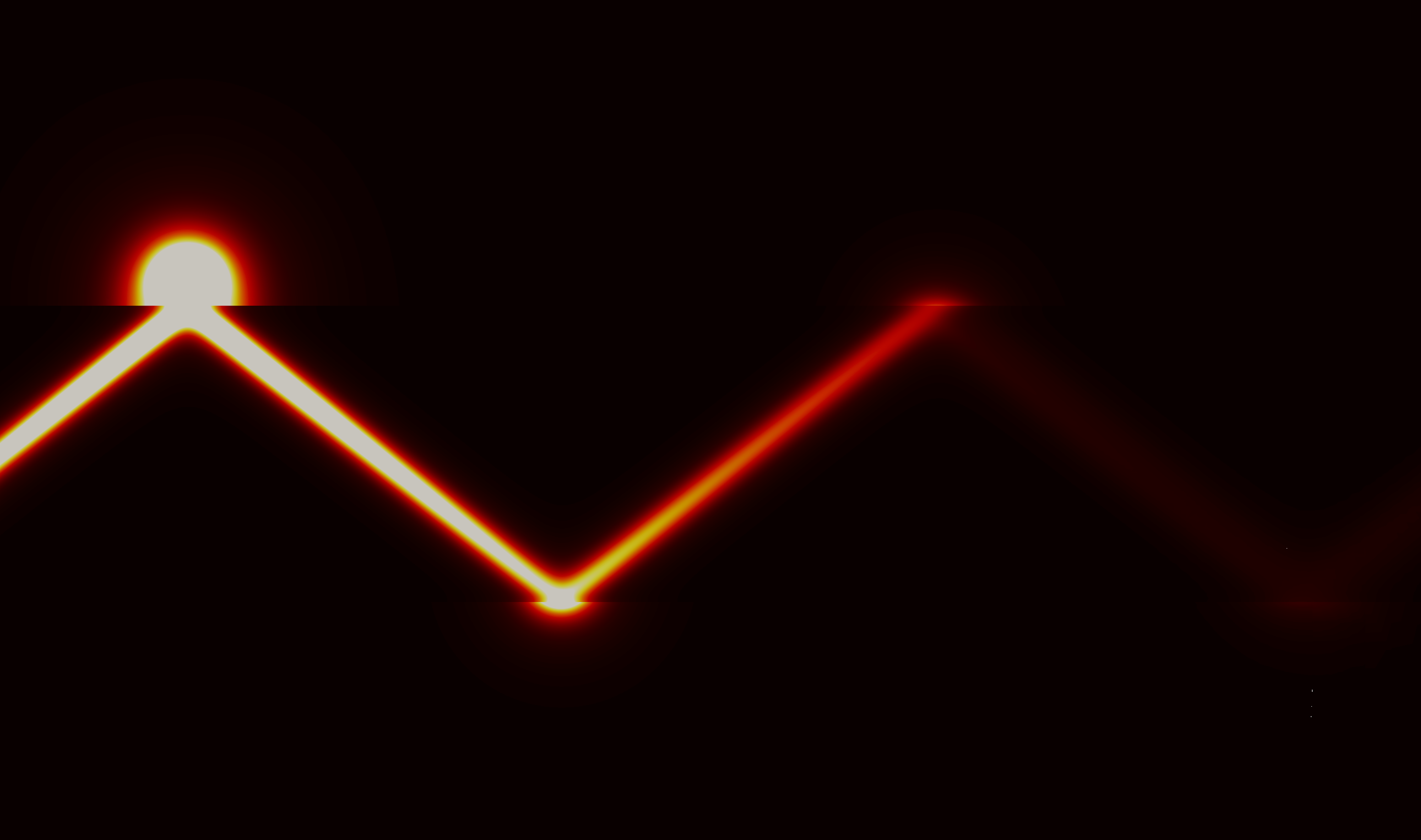}};
    \draw[dashed, white] (0.2,-0.6)--(0.2,0.4);
    \draw[dashed, white] (2.05,-0.6)--(2.05,0.4);
  \end{tikzpicture}\quad}
  \subfloat[]{\quad
  \begin{tikzpicture}
    \node at (0,0)
    {\includegraphics[width=0.42\textwidth]{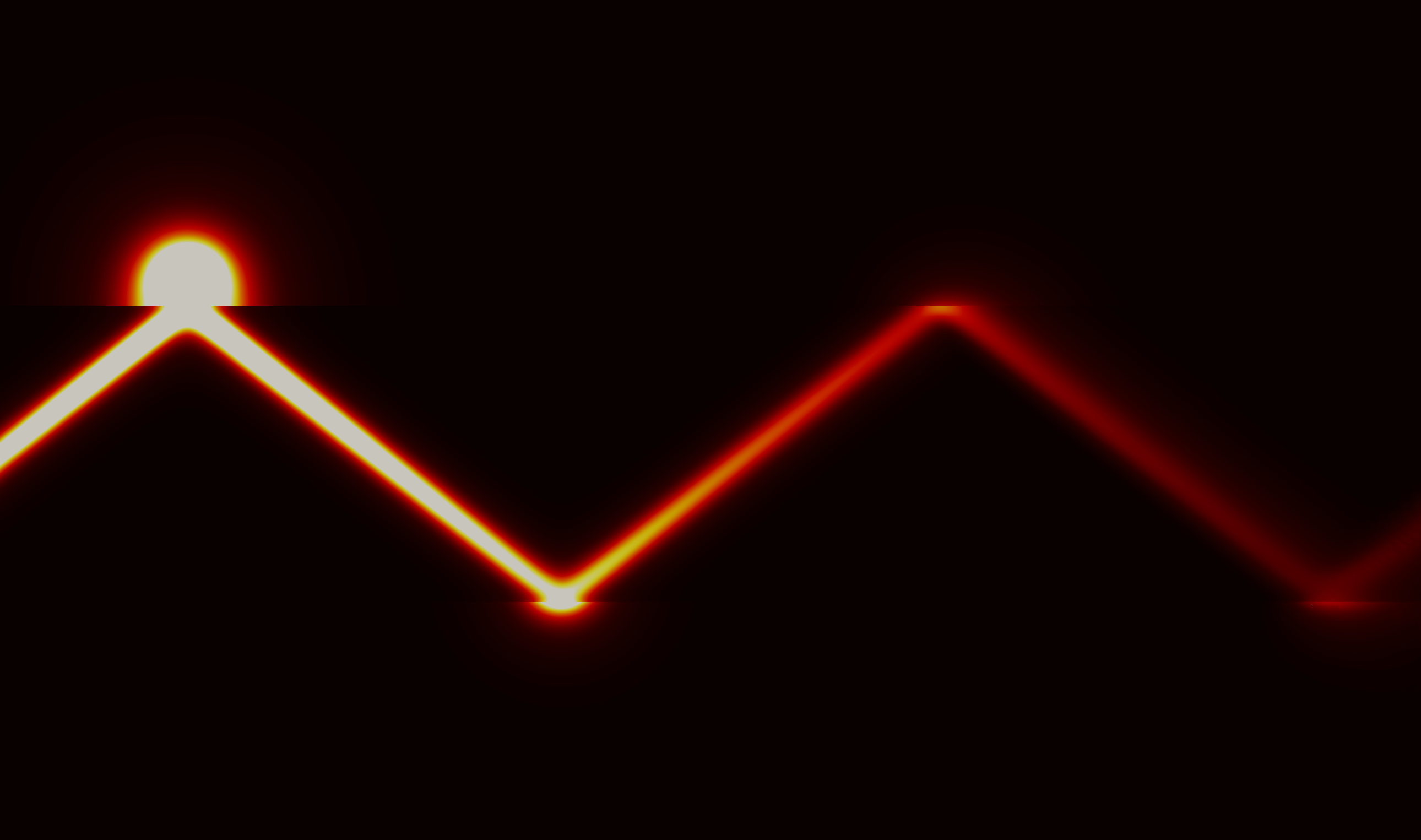}};
    \draw[dashed, white] (0.2,-0.6)--(0.2,0.4);
    \draw[dashed, white] (2.05,-0.6)--(2.05,0.4);
  \end{tikzpicture}\quad}

  \subfloat[]{\quad
    \begin{tikzpicture}
      \node at (0,0)
      {\includegraphics[width=0.42\textwidth]{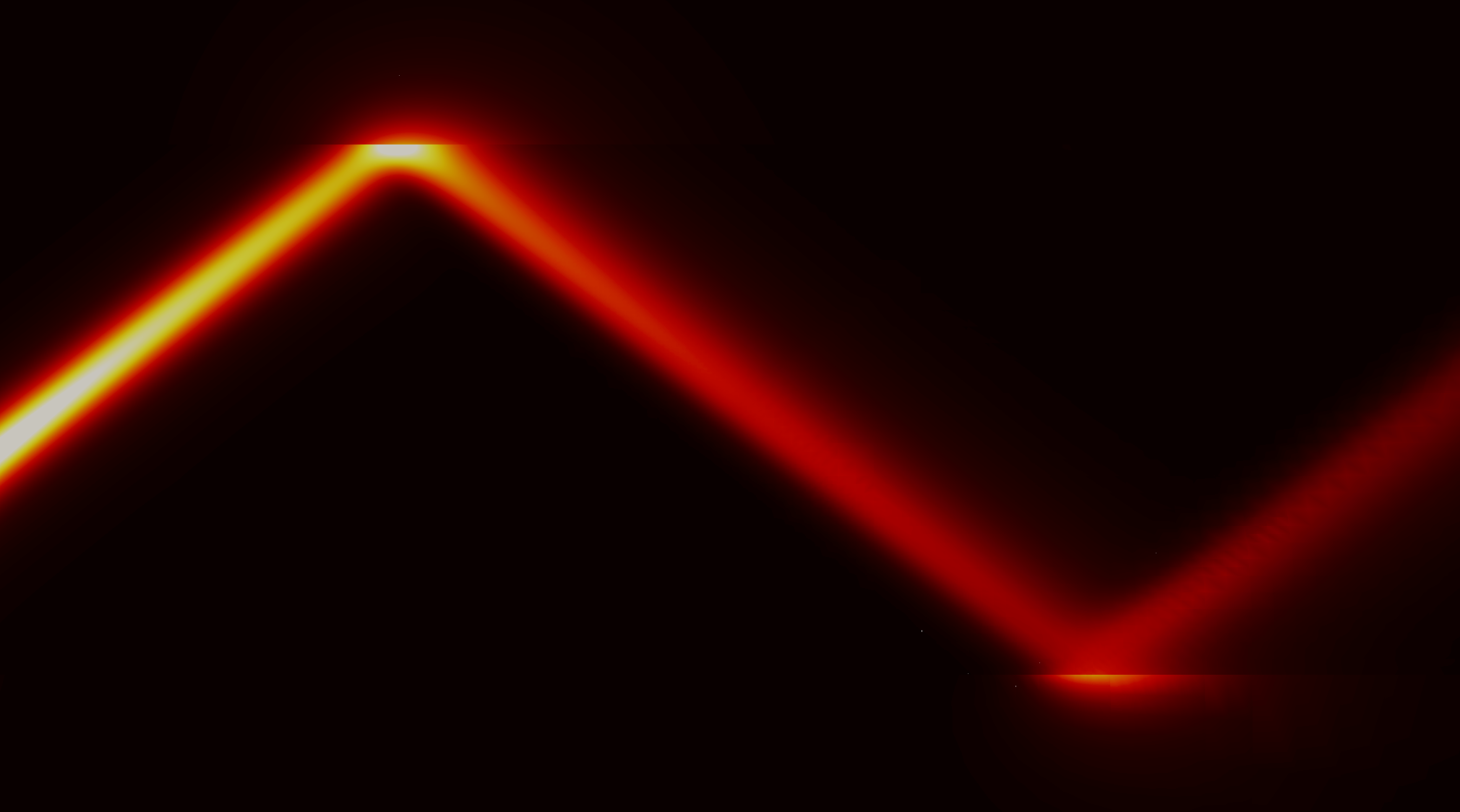}};
      \draw[white, thick] (1.80,-1.3)--(-1.37,+1.2);
      \draw[dashed, white, thick] (1.51,-1.3)--(-1.65,+1.2);
      \draw[white, thick] (1.51,-1.5)--(1.51,-1.8);
      \draw[white, thick] (1.80,-1.5)--(1.80,-1.8);
      \draw[white, thick] (1.51,-1.65)--(1.80,-1.65);
    \end{tikzpicture}\quad}

  \caption{Graphene-hBN modulator without a dielectric gap.
    (a)
    Transmission as a function of chemical potential of graphene, $\mu$. The dashed lines correspond to a transmission
    ($t$ = -3.88 dB, -3.09 dB, -294 dB) taking only losses in hBN into
    account (a geometry without graphene).
    (b) Lateral shift experienced by a hyperbolic beam hitting the
    graphene-hBN interface (for $\omega=1500 \text{cm}^{-1}$). (c,d)
    Distribution of the electric field, $\mathbf{E}$, inside the system for
    $\mu = 0.1$ eV (c) and 0.3 eV (d). The white
    dashed lines indicate the measurement positions for the field
    intensity used for the transmission calculation. (e) Lateral shift in
    the position of the maximum of the beam field intensity on the hBN-substrate interface. The
    dashed line shows the beam position in a configuration without
    graphene.}
  \label{Figure2}
  % (a)
  % frequency  comparison  minimal   maximal
  % 1450 cm-1  -3.88 dB    -6.27 dB -3.65 dB
  % 1500 cm-1  -3.09 dB    -7.83 dB -2.89 dB
  % 1550 cm-1  -2.94 dB    -9.39 dB -2.50 dB
\end{figure}
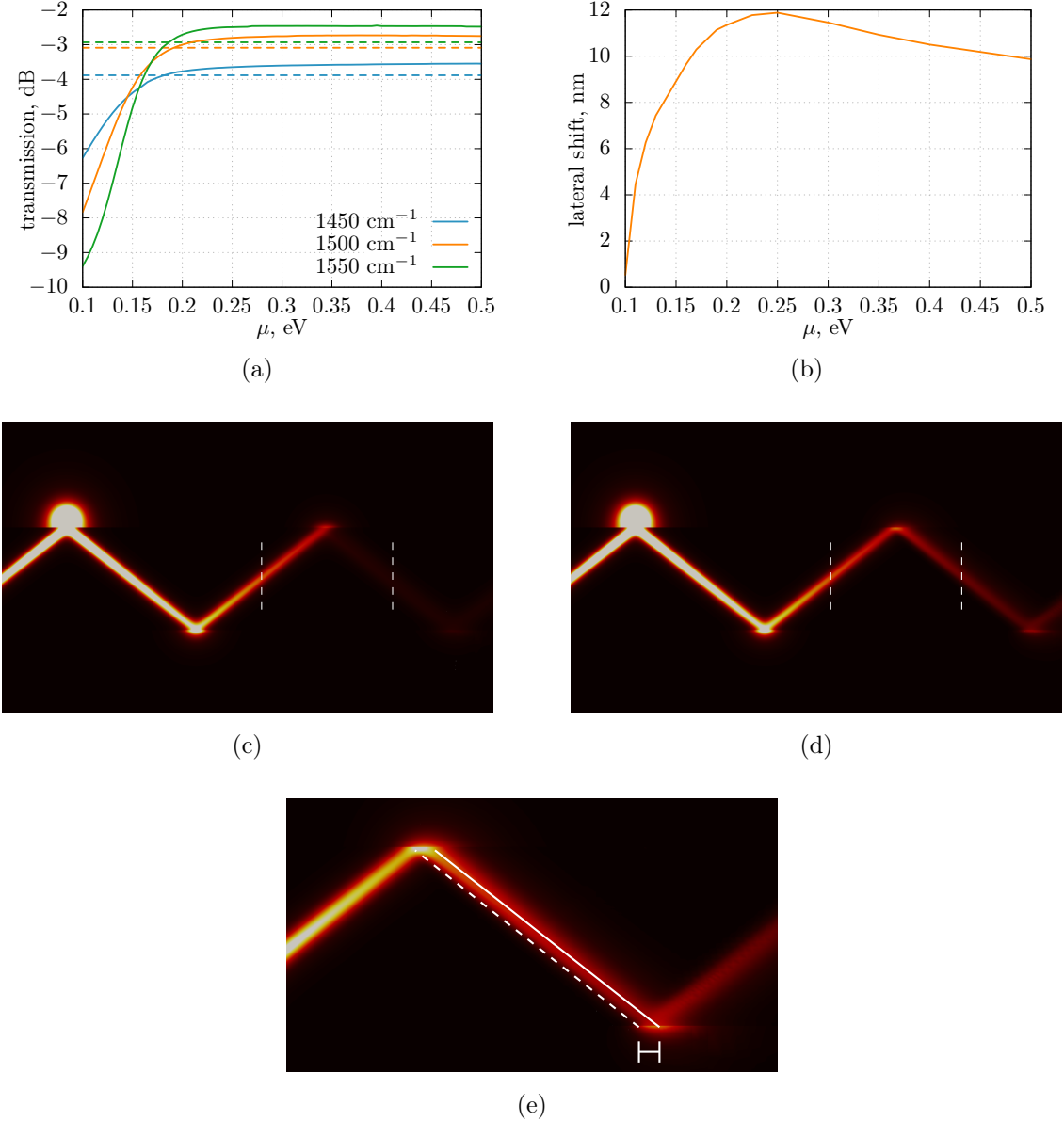

We begin by numerically examining the performance of a modulator based on a graphene-hBN system without dielectric gap; see Fig.~\ref{Figure2}. All computations are preformed with a
time-harmonic Maxwell solver \cite{maier2017} based on the finite element
toolkit deal.II \cite{dealii85}. A hyperbolic ray, excited at the upper
edge of the hBN slab by a current line source, experiences reflection from
the bottom of the hBN slab before reaching the graphene-hBN interface on
top of the slab where electro-optical modulation occurs. The ray excites
hybridized modes at the graphene-hBN surface (Figs.~\ref{Figure1}f, g) which
propagate along the interface until re-radiating their energy back to hBN
hyperbolic modes. Losses experienced by the surface mode are responsible
for the amplitude modulation of the hyperbolic signal
(Fig.~\ref{Figure2}a). The apparent lateral shift of the "reflected"
hyperbolic ray with respect to the incident hyperbolic ray (Fig.
\ref{Figure2}b) is a manifistation of the Goos-Hanchen effect
\cite{Fogler2015}, which we elaborate further below.

We define a transmission value by taking the quotient of the beam intensity
around the center of the hBN slab directly before it reaches and directly
after it decouples from the graphene-hBN interface (see white dashed lines
in Figs.~\ref{Figure2}c, d). The simulations are
done for 1450 - 1550 $\text{cm}^{-1}$ and $\mu=$0.1 -- 0.5 eV. For values
$\mu<0.1$ eV the imaginary part $\text{Im}(\sigma)$ of the graphene optical
conductivity as given by Eq. \eqref{Eq:graphene_conductivity} becomes negative
and graphene cannot support TM plasmon modes anymore. As one can see the
transmission increases monotonously from a chemical potential of 0.1 eV to
around 0.25 eV. In the case of the graphene-hBN modulator
without dielectric gap, the modulation depth achieved varies between 2.62 dB
(for 1450 cm$^{-1}$) and 6.89 dB (for 1550 cm$^{-1}$) (see Fig.
\ref{Figure2}a).

The transmission plateaus for values of chemical potential above 0.25 eV as
graphene-hBN plasmon-polariton losses reach a minimal value (see
Fig.~\ref{Figure1}g). It should be pointed out that the transmission of the
system in the "on" state (i.e. $\mu \approx 0.25$ eV) is only marginally lower
than the transmission of hBN without graphene being present, as at high
enough chemical potential the losses are mostly defined by phonons in hBN.
Despite the relatively low losses in the "on" state, the performance of the modulator (without dielectric gap) suffers from quite high transmission in the "off" state (i.e.
for $\mu = 0.1$ eV). This is due to the fact that surface
plasmon-polaritons launched at the interface by hyperbolic rays travel for
a relatively short distance before being radiatively decoupled back into
bulk hyperbolic modes of hBN (see Figs. \ref{Figure2}b, e). Estimations of plasmon travel distance based on the lateral shift experienced by hyperbolic ray hitting graphene-hBN interface are presented in Fig. \ref{Figure2}b.
We measured this lateral shift by computing the displacement of the hyperbolic beam maximum on the hBN-substrate interface
from a reference position obtained with a computation without graphene; see
Fig.~\ref{Figure2}e. As one can see, the value of lateral shift increases from around 2 nm for 0.1 eV to 12 nm at 0.25 eV.

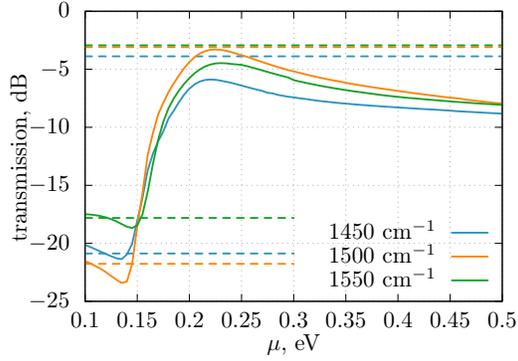
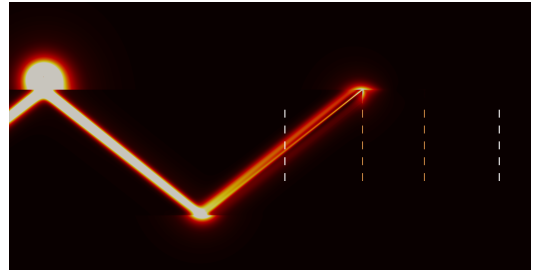
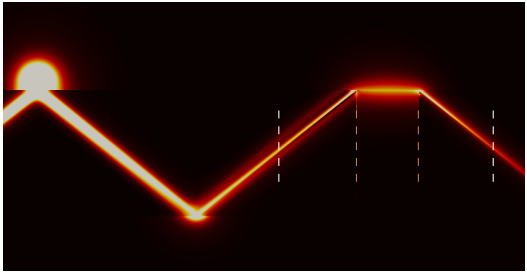
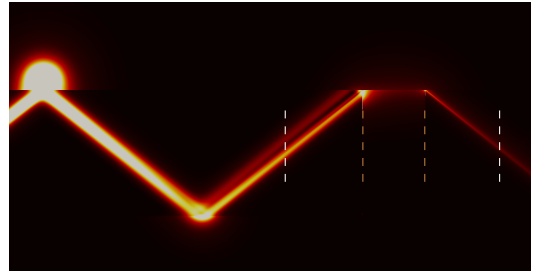
\begin{figure}[tbp]
  \centering
  \subfloat[]{\quad\includegraphics[width=0.42\textwidth]{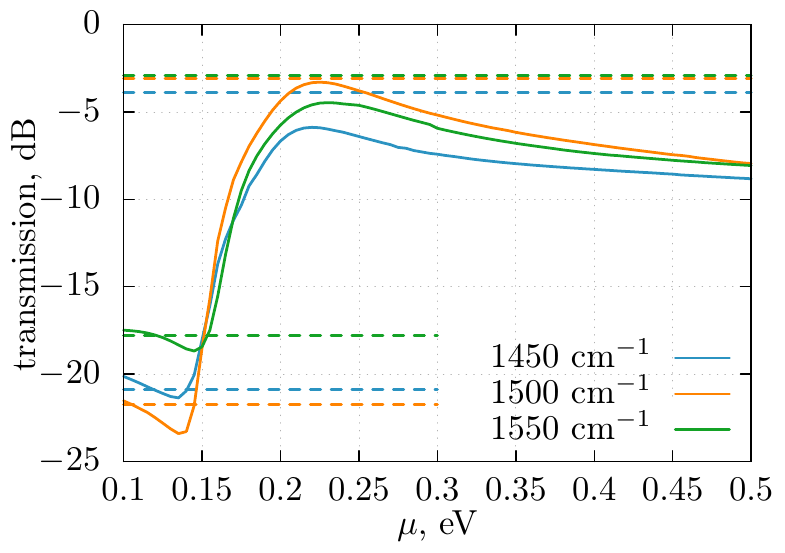}\quad}
  \subfloat[]{\quad
  \begin{tikzpicture}
    \node at (0,0)
      {\includegraphics[width=0.42\textwidth]{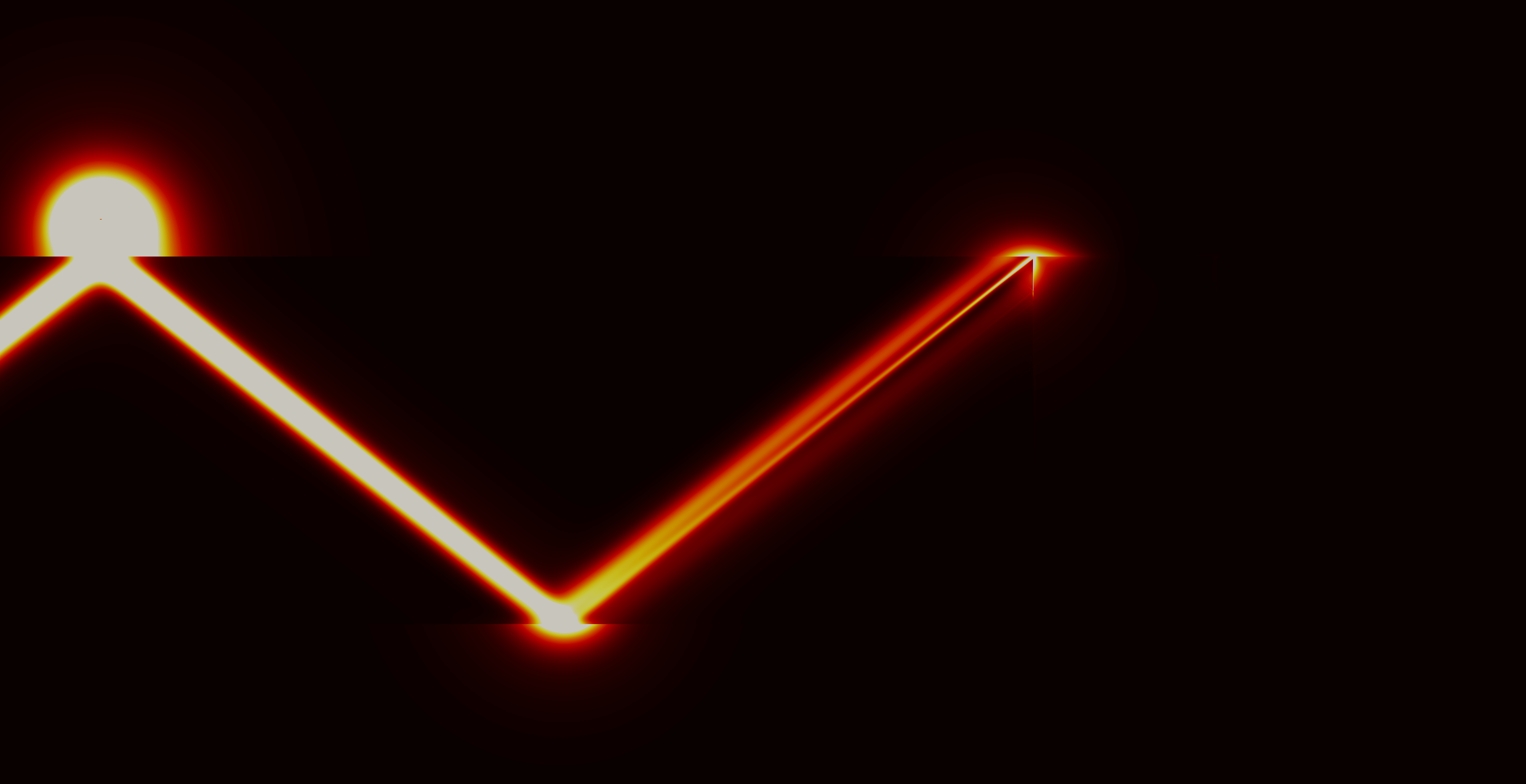}};
    \draw[dashed, white] (0.2,-0.6)--(0.2,0.4);
    \draw[dashed, brown] (1.232,-0.6)--(1.232,0.4);
    \draw[dashed, brown] (2.055,-0.6)--(2.055,0.4);
    \draw[dashed, white] (3.05,-0.6)--(3.05,0.4);
  \end{tikzpicture}\quad}

  \subfloat[]{\quad
  \begin{tikzpicture}
    \node at (0,0)
      {\includegraphics[width=0.42\textwidth]{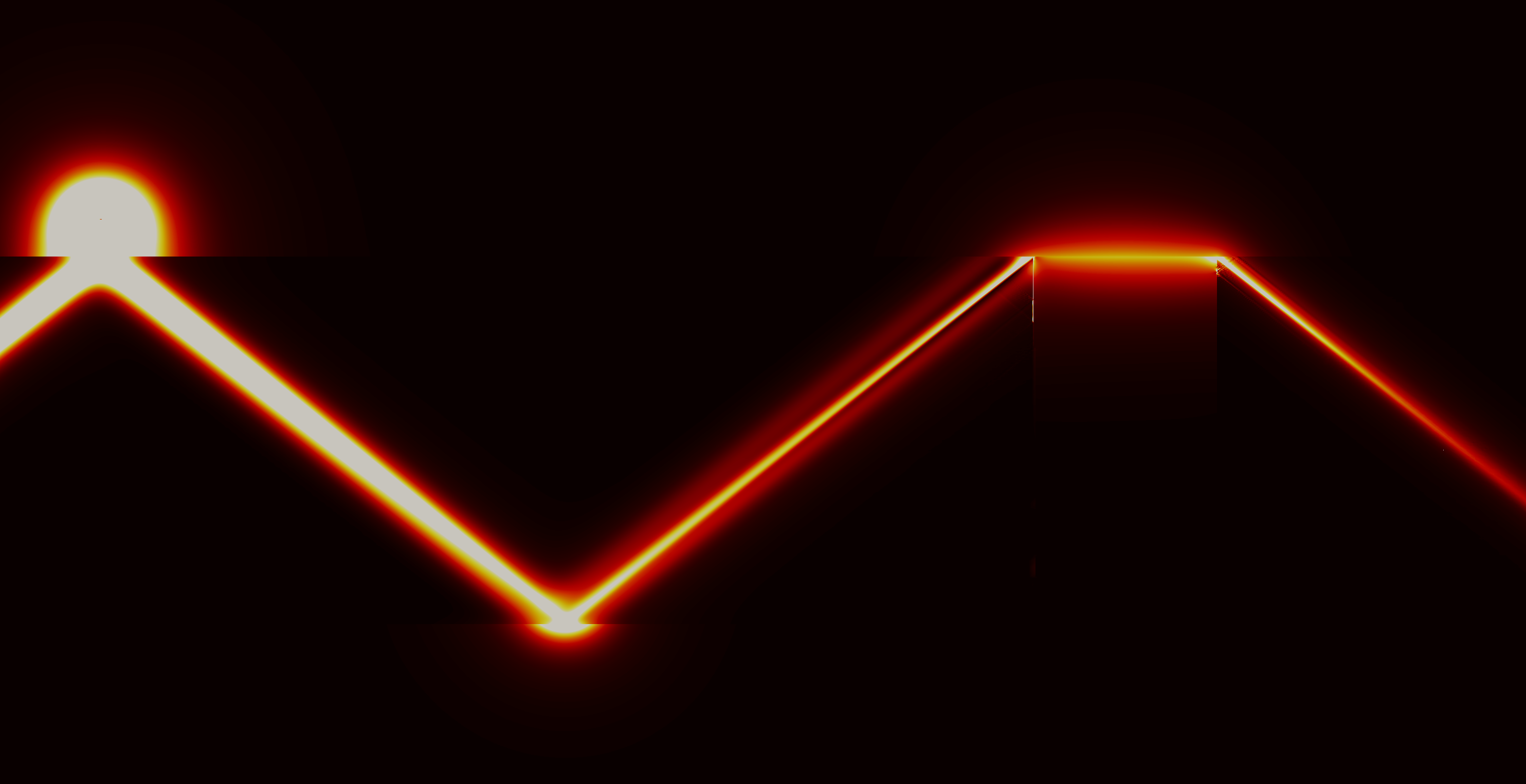}};
    \draw[dashed, white] (0.2,-0.6)--(0.2,0.4);
    \draw[dashed, brown] (1.232,-0.6)--(1.232,0.4);
    \draw[dashed, brown] (2.055,-0.6)--(2.055,0.4);
    \draw[dashed, white] (3.05,-0.6)--(3.05,0.4);
  \end{tikzpicture}\quad}
  \subfloat[]{\quad
  \begin{tikzpicture}
    \node at (0,0)
      {\includegraphics[width=0.42\textwidth]{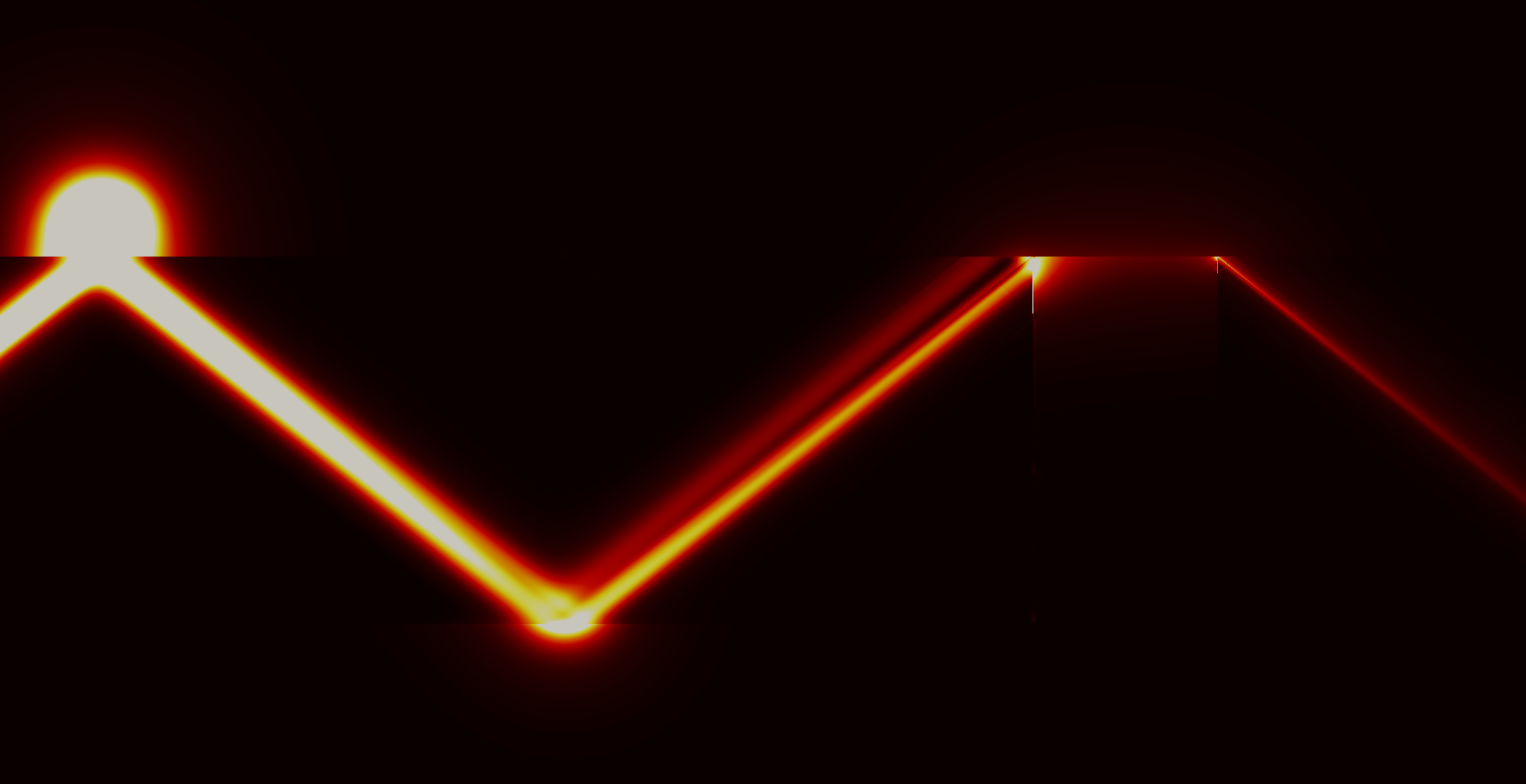}};
    \draw[dashed, white] (0.2,-0.6)--(0.2,0.4);
    \draw[dashed, brown] (1.232,-0.6)--(1.232,0.4);
    \draw[dashed, brown] (2.055,-0.6)--(2.055,0.4);
    \draw[dashed, white] (3.05,-0.6)--(3.05,0.4);
  \end{tikzpicture}\quad}

  \caption{Graphene-hBN modulator with a dielectric gap of width $w = 50$
    nm. (a) Transmission as a function of chemical potential of graphene,
    $\mu$. The lower dashed lines are transmission values ($t$ = -20.9 dB, -21.8
    dB, -17.8 dB) for a geometry with gap but without graphene; the upper
    dashed lines are the transmission ($t$ = -3.88 dB, -3.09 dB, -2.94 dB)
    taking only losses in hBN into account (a geometry without graphene and
    without gap)
    (b, c, d) Distribution of the electric field, $\mathbf{E}$, inside the
    system for $\omega=1500$ cm$^{-1}$ and for $\mu = 0.13$ eV (b), 0.22
    eV (c) and 0.4 eV (d). The white dashed lines indicate to the
    measurement positions for the field intensity used for the transmission
    calculation.}
  \label{Figure3}
\end{figure}

In order to increase the losses in the "off" state, the distance traveled
by interface plasmon-polaritons before they radiatively decouple back into hBN has
to be increased. Our proposal consists of carving out a length of the hBN
slab as illustrated in Fig. 1a. The hyperbolic ray launches surface
plasmons on the graphene surface above the gap. These surface plasmons will
not be able to decouple into phonon-polaritons of the hBN slab until they
reach the other side of the gap. Thus, the
losses experienced by plasmons in the "off" state can be significantly
increased by increasing the gap width $w$. Unless specified otherwise, we
assume that the material inside the gap is air, i.e. the gap permittivity
is $\varepsilon_g = 1$. The losses in graphene suspended above the gap are
presented in Fig.~\ref{Figure1}e.

The results of the transmission calculation are shown in
Fig.~\ref{Figure3}a. The transmission increases from a minimum value of around -20 dB at 0.13 eV
to -(3--7) dB at around 0.22eV. We obtain an overall modulation depth
ranging between 14.2 dB (for 1550 cm$^{-1}$) to 20.1 dB (for 1500
cm$^{-1}$). Thus, we can see that the proposed modulator geometry provides efficient modulation in the broad frequency range coinciding with the type-II hyperbolicity region of hBN.

A qualitative understanding of such a behavior can be obtained by studying
Fig.~\ref{Figure1}e and Figs.~\ref{Figure3}b-d. First of all, we observed
that within a range of chemical potential values (around 0.13 eV, see also Fig. \ref{Figure3}b), the
transmission across the dielectric gap covered by graphene is less than the transmission in case when
graphene is absent. Clearly, even if there are no surface plasmons to mediate coupling across
the gap between hBN slabs, the edge of the slab where the hyperbolic ray hits the hBN interface serves as a source (though inefficient) of the secondary hyperbolic polaritons in the hBN slab across the gap
(transmission of order of -20 dB). However, when the graphene is present, the energy of the hyperbolic ray is
mostly transferred to surface modes of the graphene, rather than to the
electric field at the edge. As the graphene surface plasmons decay fast
(see Fig.~\ref{Figure3}b) when $\mu \approx 0.13$ eV, the amount of energy  delivered to the adjacent hBN slab is very small which leads to
a strong dip in transmission. As the chemical potential increases plasmon
losses in graphene decrease (see Fig.~\ref{Figure1}e) which leads to an
increase in the plasmon propagation distance (see Fig.~\ref{Figure3}b) and
thus to an increased transmission across the gap. The plasmon losses,
however, are a monotonously decreasing function whereas the transmission
across the gap has pronounced peak around 0.22 eV with transmission values
close to those of a system without a gap.

\begin{figure}[tbp]
  \centering
  \subfloat[]{\quad\includegraphics[width=0.42\textwidth]{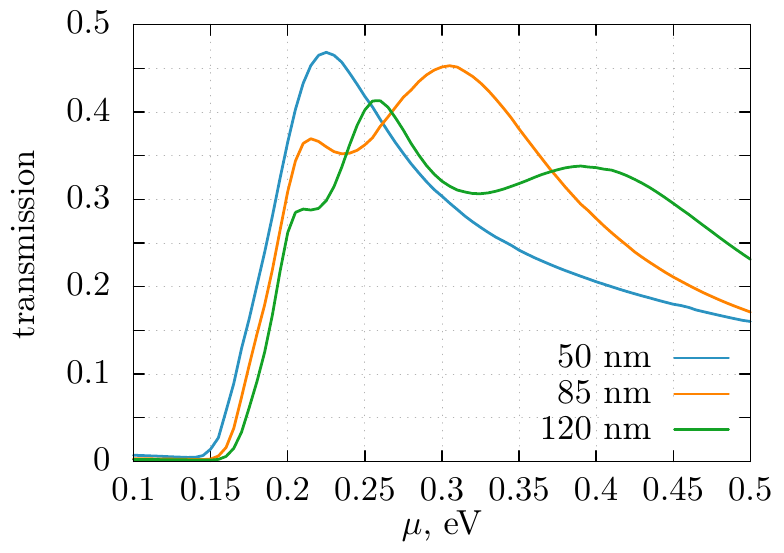}\quad}
  \subfloat[]{\quad\includegraphics[width=0.42\textwidth]{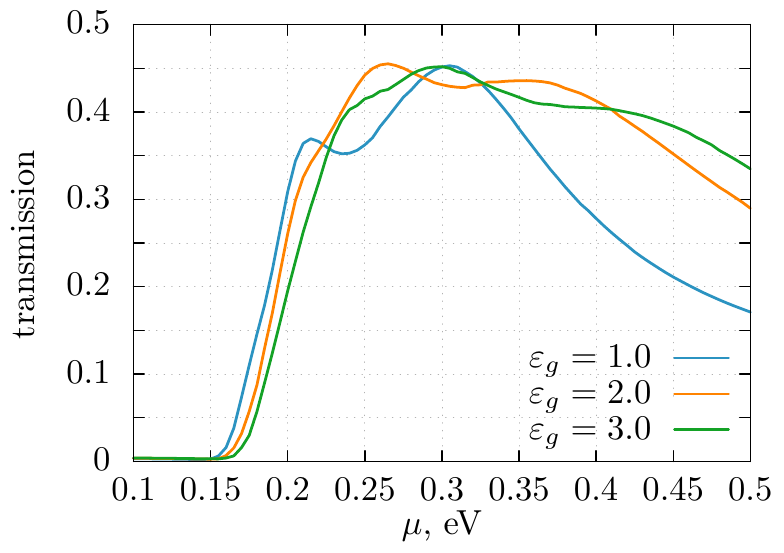}\quad}

  \subfloat[]{\quad\includegraphics[width=0.42\textwidth]{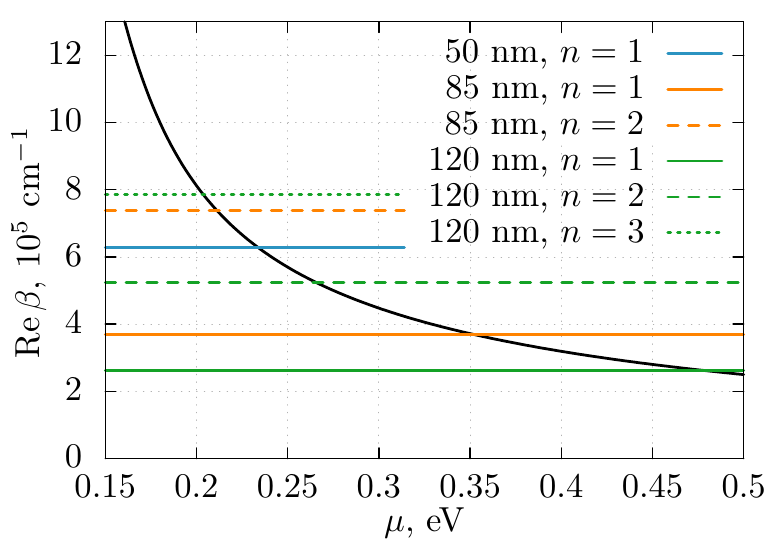}\quad}
  \subfloat[]{\quad\includegraphics[width=0.42\textwidth]{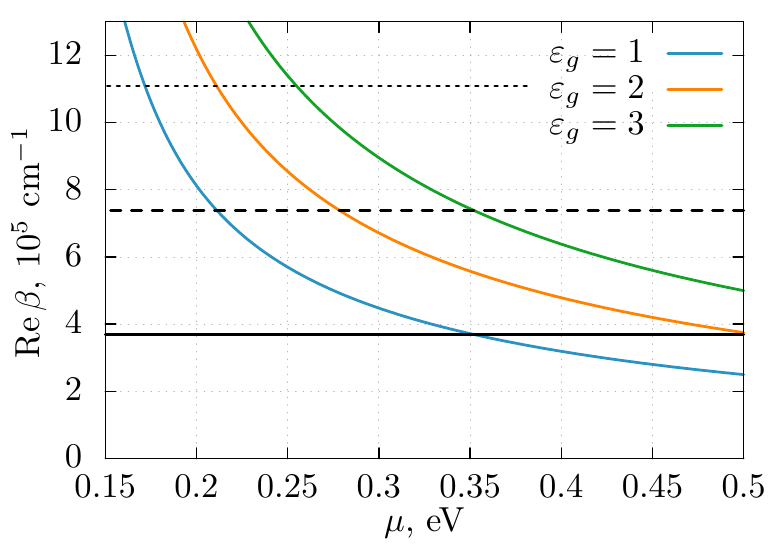}\quad}

  \caption{(a,b) Transmission across a dielectric filled gap as a function
    of chemical potential $\mu$ for (a) different gap widths, $w$, assuming
    that gap material is vacuum, $\varepsilon_g = 1$, and (b) different gap
    materials, assuming that gap width is 85 nm. (c) Dispersion relation
    for plasmons in graphene suspended in vacuum (black line). Horizontal
    lines depict wavenumbers of geometrical resonances of surface plasmons
    (see Eq. \eqref{Eq:reson}).  Solid horizontal lines correspond to $n=1$,
    dashed - $n = 2$, dashed-dotted - $n = 3$. (d) Dispersion relation for
    plasmons in graphene on dielectric. Horizontal black lines depict wavenumbers
    of geometrical resonances of surface plasmons (see Eq.
    \eqref{Eq:reson}) in the cavity of width 85 nm, where solid line
    corresponds to $n=1$, dashed - $n = 2$, dashed-dotted - $n = 3$.}
  \label{Figure4}
\end{figure}
In order to understand this behavior we should take into account that graphene
suspended above the dielectric gap behaves effectively
as an electrostatically defined ribbon (i.e., a Fabri-Perot
resonator), with gap walls serving as ribbon edges (i.e. the ribbon width is
equal to that of the gap width). The origin of the pronounced peak in the
transmission are geometrical resonances of surface plasmons in the "ribbon",
defined by the condition
\begin{equation}
  \label{Eq:reson}
  \beta w = n \pi,
\end{equation}
where $n = 1,2,3, \ldots$ defines the resonance order. A strong increase of
the electric field for values of the chemical potential corresponding to
the geometrical resonance causes a strong increase in transmission, while
the case of a weak off-resonant electric field in the cavity leads to a low
transmission. 

In order to corroborate this conclusion we calculated
transmission across gaps of different widths assuming that the gap material
is vacuum (see Fig.~\ref{Figure4}a), and across a gap of fixed width, $w =
85$ nm, assuming that the gap is filled with different dielectrics (see
Fig.~\ref{Figure4}b). With the increase of gap width, the resonance peak
shift towards higher values of chemical potential. Fig.~\ref{Figure4}c
shows the dispersion relation for a surface plasmon in vacuum superimposed
on a set of horizontal lines depicting resonant wavenumbers of cavity
modes, obtained using Eq. \eqref{Eq:reson}. It follows that the shift is due
to a decrease of plasmon wavenumber with the increase in chemical
potential. As the increase of the cavity width causes a decrease of the
resonance wavenumber, the resonance condition defined by Eq.
\eqref{Eq:reson} requires a simultaneous increase in chemical potential. It
should be pointed out that the number of resonance peaks in the
transmission spectra increases with the increase in cavity width (two
resonances for $w = 85$ nm and three resonances for $w = 120$ nm), as a
wider cavity can support high-order resonant modes ($n = 2$ and $n = 3$).

A fast increase of the plasmon wavenumber at low values of the chemical
potential leads to the resonance in transmission spectra around 0.22 eV, the position of which is practically not affected by gap width (compare condition
for $n = 2$ resonance for $w = 85$ nm and $n = 3$ resonance for $w = 120$
nm in Fig. 4c). The resonance around 0.22 eV is very narrow as plasmonic
losses increase drastically below 0.2 eV (see Fig.~\ref{Figure1}e) and the
resonance becomes overdamped. As the chemical
potential increases the wavelength of surface plasmons in graphene,
$\lambda_{pl} = 2\pi/\mathrm{Re} \beta$, increases (see Fig.
\ref{Figure1}d) and graphene plasmons become off-resonant when
$\lambda_{pl} \gg w$. This causes a strong decrease in the transmission
at high values of the chemical potential (see Figs. \ref{Figure4}a,b and SI
for details).

The analysis of the gap transmission for different dielectric materials
filling the gap (Figs.~\ref{Figure4}b, d) supports the conclusion that the
observed transmission peaks are of resonant nature. Again, we can observe a
shift of the peaks in the transmission spectra with increasing values of
the cavity dielectric function towards higher values of chemical potential
(see Fig.~\ref{Figure4}b). This is due to the increase in plasmon wavenumber
with increasing relative permittivity of the substrate
(see Fig.~\ref{Figure4}d) caused by a stronger
electrostatic screening.

Concluding, in the paper we proposed the concept of an ultra-compact
broadband graphene-hBN modulator operating in the frequency range of
type-II hyperbolicity of hBN (1450 cm$^{-1}$ -- 1500 cm$^{-1}$). The
modulating region consists of a few tens of nanometers wide graphene
covered gap carved within the hBN slab. By electrostatically gating
the graphene, we can turn the coupling between hBN hyperbolic modes on each
side of the gap on and off, thus providing efficient modulation of the
signal transferred across the gap. Such a modulator has insertion losses as
low as 3 dB and a modulation depth as high as 20 dB.

\begin{acknowledgement}
MM, AN, TL, and ML have been partially supported by the Army Research
Office (ARO) Multidisciplinary University Research Initiative (MURI) Award
No. W911NF-14-1-0247.
\end{acknowledgement}

\begin{suppinfo}
	Relation between chemical potential and mobility in
	graphene; derivation of Eq. \eqref{Eq:surface_hBN-graphene} for surface
	polaritons in air-graphene-hBN-substrate system; dispersion of second and
	third-order modes in air-graphene-hBN-substrate system; dependence of transmission across the gap on the chemical potential; dependence of modulation depth and transmission in "off" and "on" state on the gap width.
\end{suppinfo}

\providecommand{\latin}[1]{#1}
\makeatletter
\providecommand{\doi}
{\begingroup\let\do\@makeother\dospecials
	\catcode`\{=1 \catcode`\}=2\doi@aux}
\providecommand{\doi@aux}[1]{\endgroup\texttt{#1}}
\makeatother
\providecommand*\mcitethebibliography{\thebibliography}
\csname @ifundefined\endcsname{endmcitethebibliography}
{\let\endmcitethebibliography\endthebibliography}{}

\end{document}